\documentclass[reprint,superscriptaddress,amsmath,amssymb,aps,prr,longbibliography]{revtex4-1}

\usepackage[utf8]{inputenc}
\usepackage{hyperref}
\hypersetup{
    colorlinks=true,
    linkcolor=blue,
    filecolor=magenta,      
    urlcolor=cyan,
}
\usepackage{graphicx}
\usepackage[caption=false]{subfig}
\usepackage{soul}
\usepackage{float}

\begin{document}

\title{Collisionless drag for a one-dimensional two-component Bose-Hubbard model}
\author{Daniele Contessi}
\affiliation{Dipartimento  di  Fisica,  Universit\`a  di  Trento,  38123  Povo,  Italy}
\author{Donato Romito}
\affiliation{INO-CNR BEC Center, 38123  Povo,  Italy}
\affiliation{Mathematical Sciences, University of Southampton, Highfield, Southampton, SO17 1BJ, United Kingdom}
\author{Matteo Rizzi}
\affiliation{Forschungszentrum J\"ulich, Institute of Quantum Control,
Peter Gr\"unberg Institut (PGI-8), 52425 J\"ulich, Germany}

\affiliation{Institute for Theoretical Physics, University of Cologne, D-50937 K\"oln, Germany}
\author{Alessio Recati}
\affiliation{INO-CNR BEC Center, 38123  Povo,  Italy}
\affiliation{Dipartimento  di  Fisica,  Universit\`a  di  Trento,  38123  Povo,  Italy}

\date{September 2020}


\begin{abstract}
We theoretically investigate the elusive Andreev-Bashkin collisionless drag for a two-component one-dimensional Bose-Hubbard model on a ring. By means of Tensor Network algorithms, we calculate superfluid stiffness matrix as a function of intra- and inter-species interactions and of the lattice filling. We then focus on the most promising region close to the so-called pair-superfluid phase, where we observe that the drag can become comparable with the total superfluid density. 
We elucidate the importance of the drag in determining the long-range behavior of the correlation functions and the spin speed of sound. In this way we are able to provide an expression for the spin Luttinger parameter $K_S$ in terms of drag and the spin susceptibility. 

Our results are promising in view of implementing the system by using ultra-cold Bose mixtures trapped in deep optical lattices, where the size of the sample is of the same order of the number of particle we simulate. 
Importantly the mesoscopicity of the system far from being detrimental appears to favour a large drag, avoiding the Berezinskii-Kosterlitz-Thouless jump at the transition to the pair superfluid phase which would reduce the region where a large drag can be observed.

\end{abstract}

\maketitle

\section{Introduction}
The dynamics of multi-component superfluids, ranging from neutron stars~\cite{Lattimer} to superconducting layers~\cite{YipDrag}, is supposed to be crucially influenced by an inter-component dissipationless drag.
Such an entrainment has been first discussed by Andreev and Bashkin in 1975 to describe the three-fluid hydrodynamics of a mixture of $^3$He and $^4$He superfluids~\cite{Andreev-Bashkin}. 
In the previous works the constitutive relations for the superfluid momenta were assumed to involve the transport of particles of one kind only~\cite{Khalatnikov57, Khalatnikov73, Gala73, Mineev}.
Andreev and Bashkin instead introduced a superfluid stiffness matrix $n^{(s)}_{\alpha\beta}$, whose off-diagonal elements make it possible that a velocity $v_\beta$ in one component generates a (super-)current $j_\alpha$ in the other component even without collisions:
\begin{equation}
    j_\alpha=\sum_\beta n^{(s)}_{\alpha\beta}v_\beta\, .
\label{eq:ABdensities}
\end{equation}
Although the Andreev-Bashkin (AB) effect should be generic of multi-component systems, it has never been directly observed (e.g., in Helium due to the very low miscibility of the two isotopes) and its dependence on the microscopic parameters has been obtained only for few cases. 

The high degree of control reached in manipulating ultra-cold multi-component Bose gases 
has sparkled the hope of having a platform for a detailed experimental study of the collisionless drag.
However, a sizeable entrainment requires relative large quantum fluctuation beyond the mean-field equation of state. In the most standard configurations, i.e., three-dimensional gases~\cite{Fil,Sudbo2009}, increasing quantum correlations amounts to increasing the interactions. Which is technically feasible, but  only partially viable, since strong interactions lead to large 3-body losses, reducing very much the life time of the gas. 
Fortunately, other routes are available to increase the role of quantum correlations, while keeping the system stable.

Very recently, configurations with reduced dimensionality have been proposed:
in particular, in~\cite{Nespolo} it has been shown that approaching the molecular phase in a double-layer dipolar gas system the drag can become increasingly large.
In~\cite{Parisi2018}, one-dimensional mixtures close to the so-called Tonks-Girardeau regime have been shown to exhibit a large entrainment.

Another possibility is to consider Hubbard-like models, by putting cold-atoms in deep optical lattices. 
In this way it is possible not only to realise strongly interacting superfluids with reduced 3-body losses, but also to study new phases that do not appear in continuous systems.
An analysis of the AB effect in a two-component single-band two-dimensional Bose-Hubbard model can be found in~\cite{Babaev2018}, where the effect of the proximity to the Mott insulating phase is discussed in detail. 

In the present work we give a detailed account of the AB drag in two-component Bose-Hubbard model on a one-dimensional ring.
The reason is many-fold.
The ring geometry is very convenient to study super-current related phenomena, both theoretically and experimentally~\cite{Kohn}. 
The presence of the lattice allows us to have a finite range of parameters in the attractive regime where the two-superfluid state is stable and the drag can be strongly enhanced~\cite{Nespolo,Babaev2018}. 
Moreover, differently to all the above mentioned works based on Quantum Monte-Carlo (QMC) techniques, here we employ a Tensor Network approach,
thus paving the way to study time dependent phenomena.

We show how finite-size effects might actually increase the visibility of the collisionless drag, by circumventing the sudden jump that characterises the phase transition to a pair-superfluid phase in the thermodynamic limit.
This is particularly relevant for typical 1D cold atomic setups, where the particle number is comparable to our numerical simulations.
In this respect, hyperfine states mixtures of $^{(39)41}$K atoms, or $^{41}$K-$^{87}$Rb mixtures (e.g.~\cite{Fattori1,Fattori2,Tarruell1,Tarruell2,Fort}), whose inter-species interaction can be tuned by exploiting Feshbach resonances, seem very promising to achieve the regimes where the elusive AB effect could be finally observed.

Furthermore, after determining the strength of the AB effect in various regimes, we use our microscopic approach to determine the susceptibility of the system and the relavant correlation function to extract the Luttinger parameter $K_S$. We show the latter satisfies a general hydrodynamic relation with the collisionless drag, a relation which is rooted in the fact that the $f$-sum rule for the spin channel is not exhausted by single-phonon excitations (see~\cite{Donato} and the Supplementary Material of~\cite{Parisi2018}). In particular, our results 
show how perturbative Luttinger liquid description of Bose-Bose Hubbard model must include an inter-velocity interaction. 



\begin{figure}[htb]
    \begin{center}
    \includegraphics[width = 0.45\textwidth]{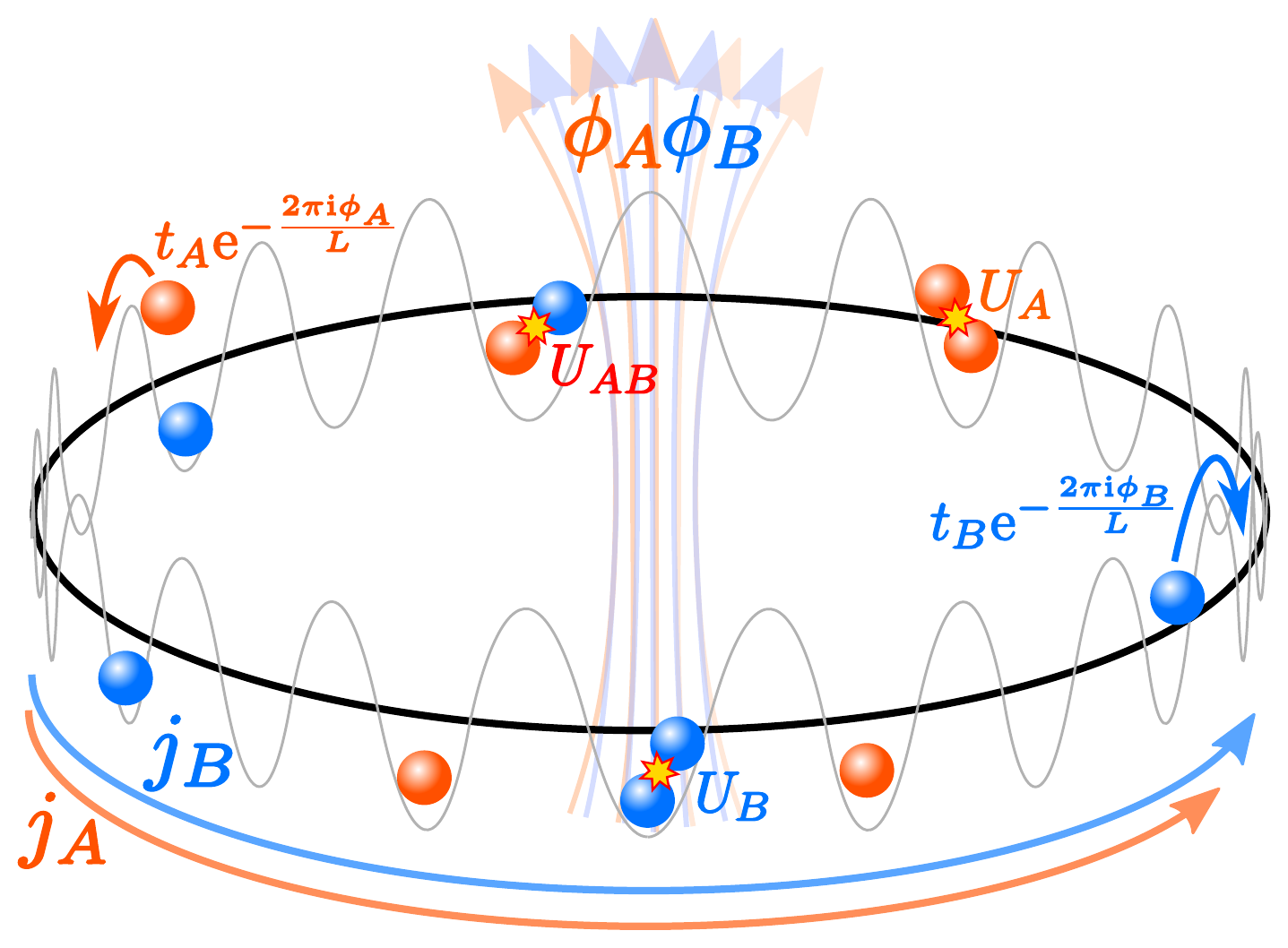}
    \end{center}
    \caption{\label{fig:sketch}Sketch of the 2-component Bose-Hubbard ring with hopping parameters $\tilde{t}_\alpha = t_\alpha e^{-i2\pi\phi_\alpha/L}$, on-site intra-species interactions $U_\alpha$ and inter-species interaction $U_{AB}$. The two fluxes $\phi_{\alpha}$ pierce the ring and give rise to bosonic currents $j_{\alpha}$ in the ground state.}
\end{figure}
\section{Model}
As sketched in Fig.~\ref{fig:sketch}, we consider a 2-species Bose-Hubbard Hamiltonian,
$H = H_A + H_B + H_{AB}$,
on a ring with $L$ sites:
\begin{eqnarray}
& & H_\alpha = \sum_{x=1}^L\left[-  \left( \tilde{t}_{\alpha} b_{x+1,\alpha}^\dagger b_{x,\alpha}^{} + \mathrm{h.c.} \right) 
    + \frac{U_\alpha}{2} n_{x,\alpha} (n_{x,\alpha} -1)\right], \nonumber \\
& & H_{AB} = U_{AB}  \sum_{x=1}^L n_{x,A} n_{x,B}\ ,
\label{eq:model}
\end{eqnarray}
where $b_{x,\alpha}^\dagger(b_{x,\alpha})$ is the bosonic creation (annihilation) operator and $n_{x,\alpha} = b_{x,\alpha}^\dagger b_{x,\alpha}$ is the number operator at site $x$ for the species $\alpha \in \{A, B\}$. Physically, the two species can be two hyperfine levels of atoms, two band indices or two different atomic elements or isotopes.
The single species Hamiltonian $H_\alpha$ accounts for the hopping between neighboring sites, with $\tilde{t}_\alpha=t_\alpha e^{-i2\pi\phi_\alpha/L}$ and $t_\alpha  ,\phi_\alpha \in\mathbb{R}^+$, and for the on-site repulsion characterised by the parameter $U_\alpha>0$.
The fluxes $\phi_\alpha$ piercing the ring are introduced in order to compute the superfluid currents and densities (see Eqs.~\eqref{eq:supercurrent}-\eqref{eq:drag}), and are equivalent to twisted periodic boundary conditions (PBC)~\cite{Leggett73}.
The Hamiltonian $H_{AB}$ describes the inter-species on-site interaction -- responsible for the collisionless drag phenomenon -- with strength $U_{AB}$.

We limit ourselves to a zero temperature, $\mathbb{Z}_2$ symmetric mixture: 
${t}_\alpha = t$, $U_\alpha = U$ and filling $\nu_\alpha \equiv N_\alpha/L = \nu/2$,
in terms of the number of atoms $N_\alpha$.
The phase diagram of the 1D model is very rich and has not yet been determined with the same accuracy as in higher dimensions~~\cite{Svistunov2003,Svistunov2004}: to our knowledge the most complete analysis can be found in~\cite{Clark}.
It is beyond the scope of the present work to bridge this gap and the full phase diagram will be reported elsewhere~\cite{PhdDaniele}.
For the purpose of the present study we are interested in only two of the possible phases: the two-superfluid (2SF) phase and the pair-superfluid (PSF) phase.
The 2SF phase is characterised by both components $A$ and $B$ being superfluid.
The low energy spectrum consists of two gapless linear (Goldstone) modes corresponding to a density (in-phase) and a spin (out-of-phase) mode.
In the PSF phase, the two components are paired and the spin-channel acquires a gap~\footnote{This phase is sometimes considered the bosonic counterpart of the much more relevant phenomenon of pair condensation occurring in fermionic systems.}. 
One of our goals here is to determine the superfluid density matrix in the 2SF phase while approaching the PSF, where the collisionless drag should saturate to its maximum value~\cite{Nespolo,Babaev2018}.

We resort to a Matrix Product States (MPS) ansatz to deal with the full many-body problem.
The model~\eqref{eq:model} is indeed not exactly solvable and our numerical treatment is an almost unbiased approach to it.
We overcome the difficulties related to PBC by employing a loop-free geometry of the tensor network and shifting the topology of the lattice into the Matrix Product Operator (MPO) representation of the Hamiltonian. The idea consists essentially in introducing nearest-neighbour couplings along a snake-like enumeration of the physical sites (see~\cite{SuppMat}).
In this way, we can reliably compute relevant quantities for systems up to $L=96$ sites 
achieving expectation values (e.g., of densities and currents) homogeneous up to $0.3\%$ along the ring.


\section{Superfluid densities and collisionless drag}

In order to determine the superfluid density matrix we need to compute the currents on the ring. As usual, the definition of the current is properly obtained through the (discrete) continuity equation:
\begin{equation}
\begin{split}
     &\frac{\partial \left\langle\hat n_{x,\alpha} (t)\right\rangle}{\partial t} = \frac{1}{\mathrm{i}\hbar}\left\langle[\hat n_{x,\alpha},\hat H]\right\rangle=\\
     &\frac{2\tilde{t}}{\hbar}\left(\mathrm{Im}\langle b_{x+1,\alpha}^\dagger b_{x,\alpha}^{} \rangle  
        - \mathrm{Im}\langle b_{x,\alpha}^\dagger b_{x-1,\alpha}^{} \rangle \right).
\end{split}
\end{equation}
Thus, one can calculate the currents using the expression  
\begin{equation}
    j_{\alpha} = \frac{2\tilde{t}}{\hbar}\ \mathrm{Im} \langle  b_{x+1,\alpha}^\dagger b_{x,\alpha}^{} \rangle 
    = \frac{1}{2\pi\hbar} \frac{\partial E}{\partial \phi_\alpha} \ ,
\label{eq:supercurrent}
\end{equation}
where the last equality has been obtained by applying the Hellmann-Feynman theorem.
%

\begin{figure}[b]
    \begin{flushleft}
    \hspace*{-0.5cm}
    \includegraphics[width = 0.50\textwidth]{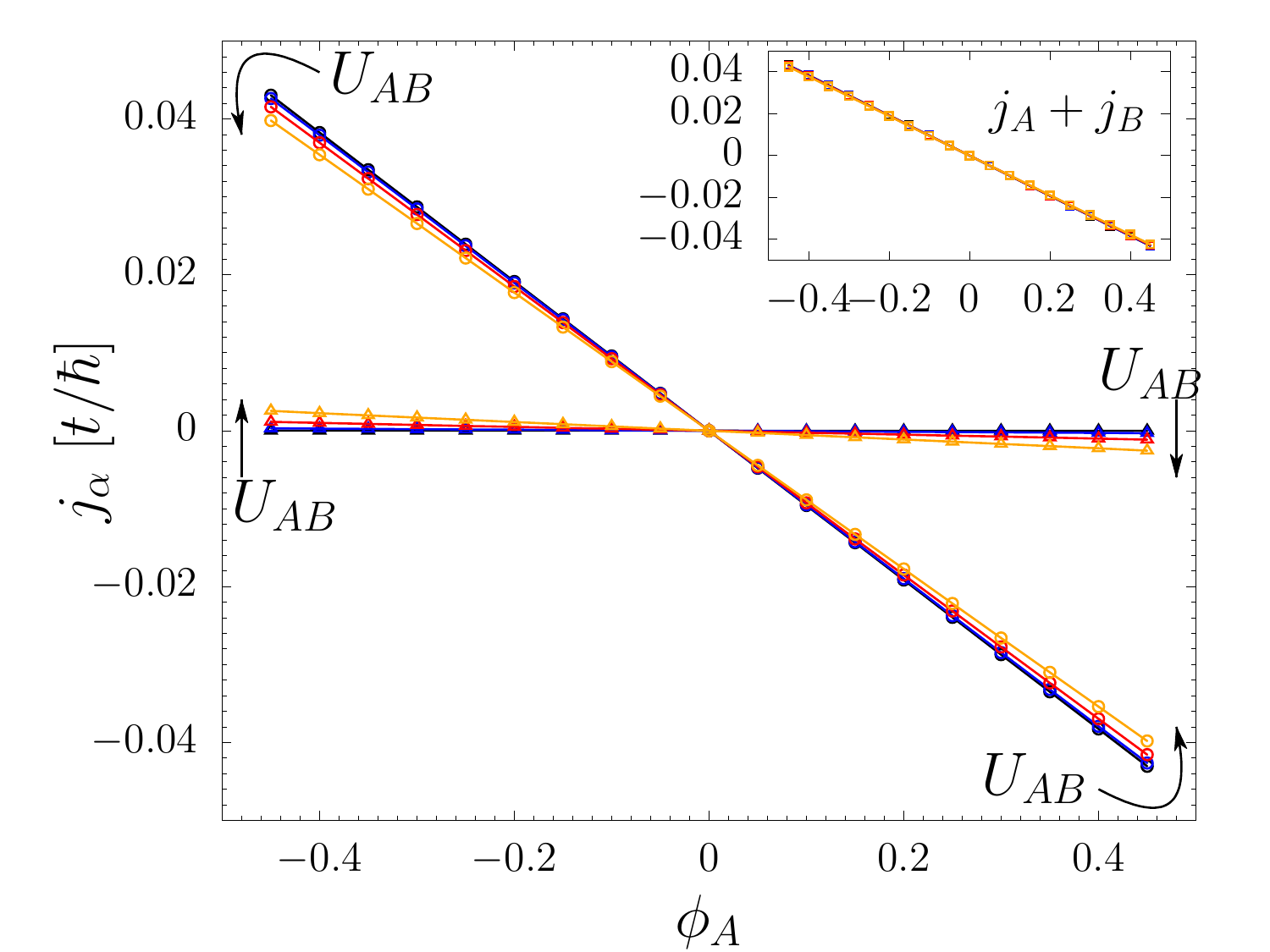}
    \caption{\label{fig:pers_curr}(color online). The superfluid currents $j_A$ and $j_B$ in presence of $\phi_A$ only for a $L=32$ system. 
    The four sets of curves (from dark to light color) correspond to $U_{AB}/U = 0.0,0.25,0.5,0.75$, with $U/t =2$ at half-filling.
    The drag density $n^{(s)}_{AB}$ is proportional to the slope of the $j_B$ curve in the limit of $\phi_A\to0$. 
    The inset demonstrates the global momentum conservation because of the constant value of the total current for all values of $U_{AB}$.
    }
    \end{flushleft}
\end{figure}
By linearising the currents for small fluxes and using the relation \begin{equation}v_\beta=\frac{\hbar}{m^*}\frac{2\pi\phi_\beta}{L},\end{equation} with  $m^*=\hbar^2/2t$ the ``band mass”, we can compute the AB~\cite{Andreev-Bashkin} superfluid density matrix $n^{(s)}_{\alpha\beta}$ of Eq.~\eqref{eq:ABdensities} as:
\begin{equation}
    n_{\alpha\beta}^{(s)} =\lim_{\phi_\alpha,\phi_\beta\to0}\frac{L m^*}{2\pi\hbar}\frac{\partial j_\alpha}{\partial \phi_\beta}= \lim_{\phi_\alpha,\phi_\beta\to0}\frac{L m^*}{(2\pi\hbar)^2} \frac{\partial^2 E}{\partial \phi_\alpha\partial\phi_\beta}.
\label{eq:drag}
\end{equation}
It is important to notice that within Tensor Network methods the total energy and current densities, computed from short-range correlations, are among the most reliable quantities to be extracted.

In Fig.~\ref{fig:pers_curr} we illustrate the effect of the AB drag on the currents:
in presence of $\phi_A$ only, the current $j_B$ is constantly zero in absence of inter-species interaction, $U_{AB} = 0$, while it increases monotonically as $U_{AB}$ increases. 
The drag density $n^{(s)}_{AB}$ is proportional to the slope of the $j_B$ curve in the limit of $\phi_A\to0$.
The plot highlights the smallness of the drag effect at a generic point in parameter space -- here the filling is $\nu=0.5$ with $U/t=2$ and $L=32$.
For completeness in the inset of Fig.~\ref{fig:pers_curr} we report the total current and we confirm that the result $j_A + j_B=2\pi\nu\phi_A/L$ is independent from the interaction.  

\subsection{Superfluid drag at half-filling}

\begin{figure}[H]
    \begin{flushleft}
    \includegraphics[width = 0.50\textwidth]{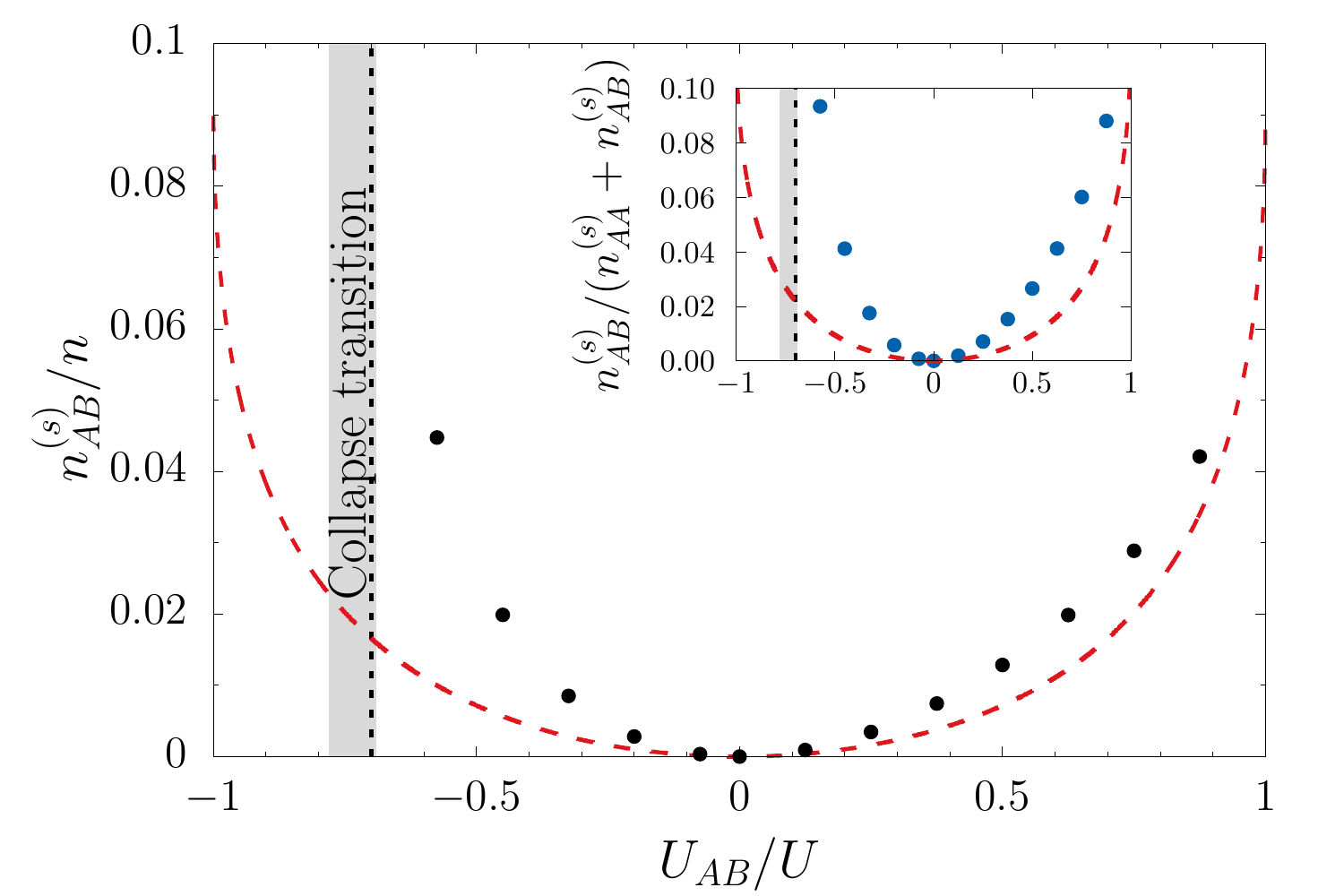}
    \caption{\label{fig:drag_half_filling}(color online).
    Superfluid drag in terms of the total density (main panel) and of the total superfluid density (inset).
    The red dashed line indicates the theoretical prediction via Bogoliubov approximation~\cite{Donato}.
    The relevant pairing correlations are responsible for the non symmetric trend exhibited by the attractive regime ($U_{AB}<0$).
}
    \end{flushleft}
\end{figure}

In Fig.~\ref{fig:drag_half_filling} we report the results for the drag density $n_{AB}^{(s)} = n_{BA}^{(s)}$ as a function of the interspecies interaction $U_{AB}/{t}$ for half total filling $\nu=0.5$ and $U/{t} = 2$. In this regime, as expected from previous analysis~\cite{Clark}, the mixture is always in the 2SF phase, until undergoes either phase separation or collapse. The location of the phase transitions is strongly dependent on the parameters of the configuration. In our case, the collapse occurs beyond the black dashed line in the shaded region.

For comparison the  Bogoliubov prediction for the entrainment is also reported. Using the method developed in Ref.~\cite{Donato} the Bogoliubov approach leads to the simple expression for the drag:
\begin{equation}
    n_{AB}^{(s)} \simeq \frac{{t}}{4L}\sum_k \frac{(\Omega_{d,k}-\Omega_{s,k})^2 k^2}{(\Omega_{d,k}+\Omega_{s,k})\Omega_{s,k}\Omega_{d,k}}\left(\frac{\sin(k)}{k}\right)^2,
\label{eq::bog}
\end{equation}
with $\Omega_{d(s),k}=\sqrt{\epsilon(k)(\epsilon(k)+2 U \nu\pm 2U_{AB} \nu)}$ the excitation energies of the density (spin) channel and $\epsilon(k)=4t\sin^2(k /2)$ the single particle dispersion relation. The sum in \eqref{eq::bog} is done on the wave vectors in the 1st Brillouin Zone $k=2\pi/L\cdot n$ with $n = 0,...,(L-1)$. 

The Bogoliubov approach turns out to be not very reliable except for very small interspecies interaction. \footnote{The same occurs in a continuous one-dimensional Bose mixture with repulsive interactions \cite{Parisi2018}, where the Monte-Carlo results are compared with the continuous version of Eq.~(\ref{eq::bog}), i.e., by replacing in Eq.
~\ref{eq::bog}: $t\rightarrow \hbar^2/2m$ and $\sin(k)/k\rightarrow 1$. }.

More importantly while Eq.~(\ref{eq::bog}) predicts a symmetric behaviour for $U_{AB}\rightarrow -U_{AB}$, the data display an evident asymmetry between the two regimes concerning both the location of the transitions and the slope of the drag increase as function of the interaction strength. In particular, the attractive mixture experiences a much steeper growth of the drag. This substantial increase can be ascribed to pairing correlations -- relevant in 1D for any value $U_{AB}<0$ -- which are not captured by the Bogoliubov approach. 

\subsection{Superfluid drag approaching the PSF phase}

\begin{figure}[h]
    \centering
    \includegraphics[width = 0.48\textwidth]{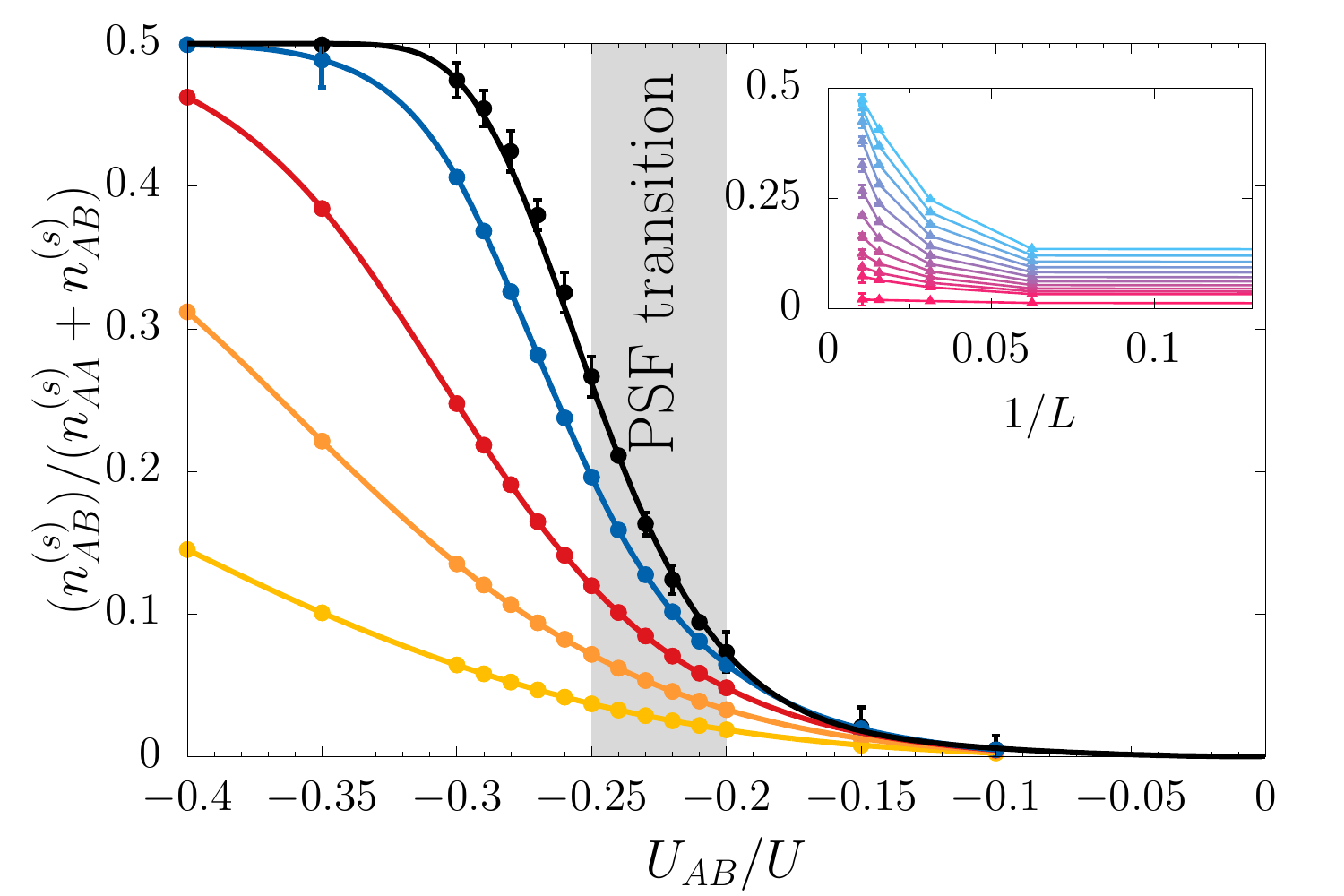}
    \caption{\label{fig:dragPSF}(color online). 
    Normalized drag with respect to the total superfluid density for different system's sizes, as a function of $U_{AB}$ for a system with $U=10t$ and $\nu=1$.
    From bottom to top (from light to dark shades) $L=8,16,32,64,96$ with corresponding bond dimensions $\chi=100,300,600,700,800$. 
    Points are data (with error bars explained in~\cite{SuppMat}), lines are artistic guide-to-the-eyes.
    The thermodynamic limit should exhibit a saturation to $(n_{AB}^{(s)})/(n_{AA}^{(s)}+n_{AB}^{(s)}) = 0.5$ in the PSF region. In the inset, the same points of the main plot are represented as a function of the inverse of the system's size $L^{-1}$ for different values of $-0.30 \leq U_{AB}/U \leq -0.15$ from top to bottom.  
    }
\end{figure}

In analogy with QMC results in the 2D case~\cite{Nespolo,Babaev2018}, we expect that the asymmetry between attractive and repulsive regimes is strongly emphasized in the regime  where a single superfluid of dimers can be reached for $U_{AB}<0$ (PSF phase). Indeed, in the latter case the flow of one component is accompanied by the flow of the other, i.e., $n_{AB}^{(s)}=n_{AA}^{(s)}$. 
The drag $n_{AB}^{(s)}$ becomes a quarter of the total superfluid density $n^{(s)} = n_{AA}^{(s)}+n_{BB}^{(s)} +2 n_{AB}^{(s)}$, hence saturates to its maximum possible value and simultaneously ceases to be interpreted as a drag coefficient. Before the saturation, the magnitude of the drag rapidly increases making the approaching to the PSF transition a very suitable region for its measurement.
The results for different system's sizes are reported in Fig.~\ref{fig:dragPSF}, where $U/t=10$ and $\nu=1$ are chosen such that the system can undergo the transition (see also \cite{Clark}). In the inset we report the behaviour of the normalized drag as a function of $L^{-1}$ for different values of the interaction. We estimate that the SF-to-PSF transition should occur in the thermodynamic limit for $U_{AB}/U\in [-0.25,-0.2]$ (shaded in region in the main panel). 
In such a limit -- belonging the transition to the Berezinskii-Kosterlitz-Thouless universality class~\cite{Clark} -- the saturation should happen as a sudden jump of the spin-superfluid density, $n_{AA}^{(s)}-n_{AB}^{(s)}$, from a finite value to zero.
We stress, however, that mesoscopic samples like the ones accessible in cold-atomic setups will display no jump, but rather a sizeable value of $n_{AB}^{(s)}$, thus making the AB collisionless drag finally observable.

\section{Collisionless Drag and Luttinger liquid parameters} 
\begin{figure*}[hbt]
\subfloat[][]{
    \includegraphics[width = 0.32\textwidth]{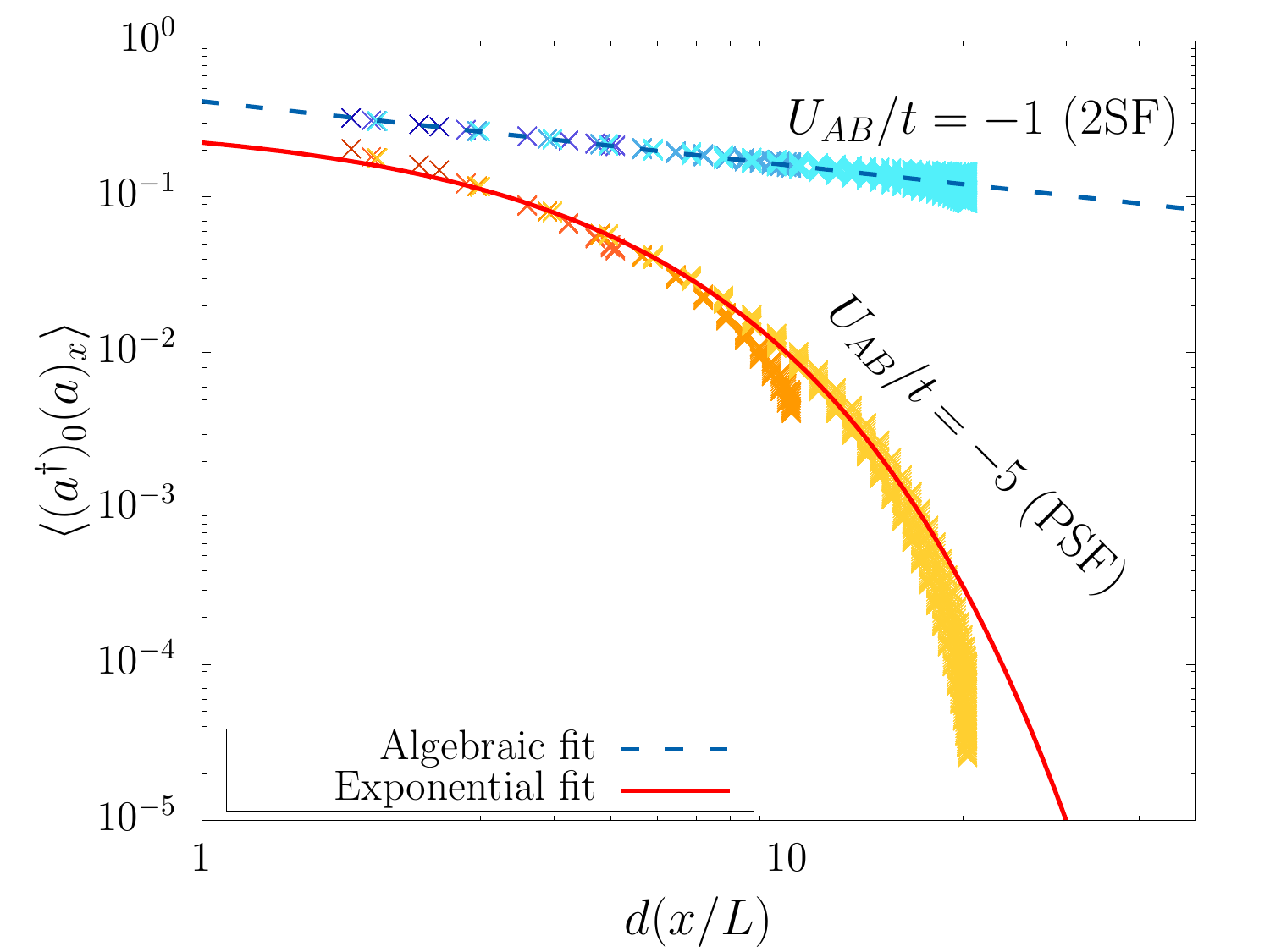}
    }
    \subfloat[][]{
    \includegraphics[width = 0.32\textwidth]{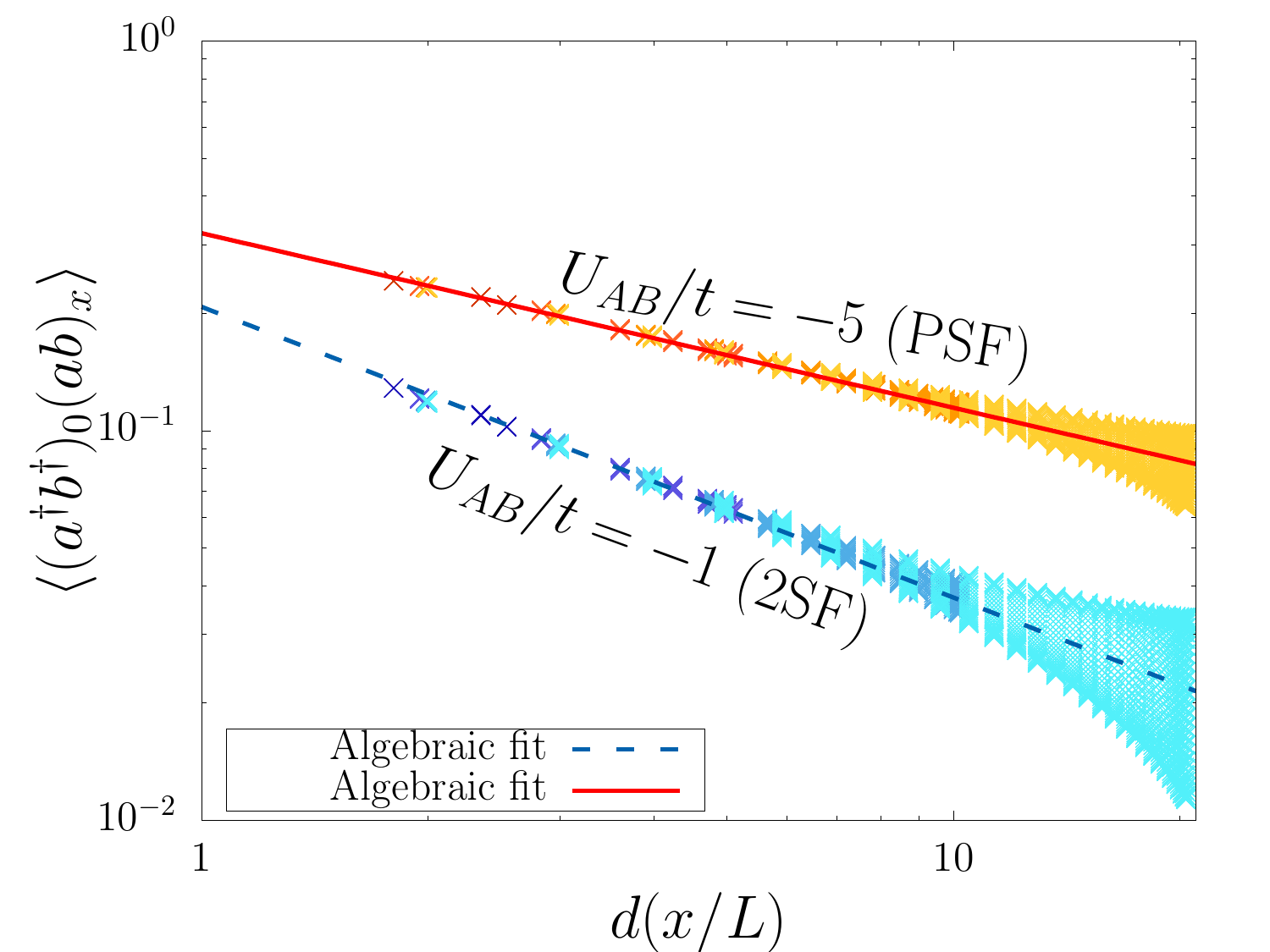}
    }
    \subfloat[][]{
    \includegraphics[width = 0.32\textwidth]{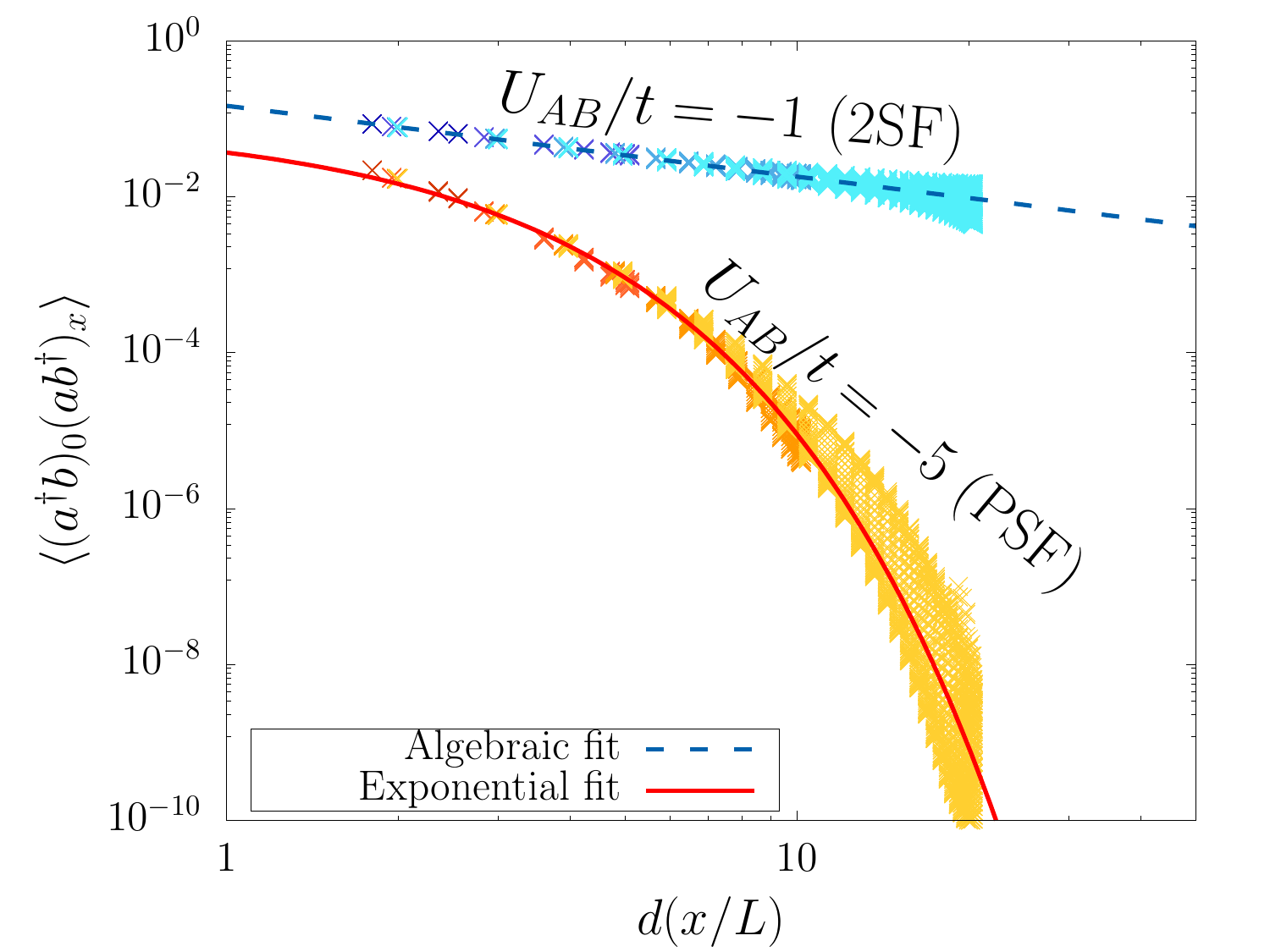}
    }
    \caption{\label{fig:correlations}(color online). The correlation functions of Eq.~\eqref{eq:lutt_param} (the (a) panel is $G_{\alpha}$, (b) is $R_{D}$ and (c) is $R_{S}$) as function of the conformal distance for a unitary total filling system $\nu=1$, $t=1$ and $U=10$. The orange points concern a regime in which the system is in PSF phase while the blue ones concern the 2SF phase and we represent with a color gradient from dark to light different system's sizes from $L=8$ to $L=64$. The dashed lines are exponential and algebraic fits depending on the expected behaviour of the functions for $U_{AB}/U = -0.1$ and the solid one for $U_{AB}/U = -0.5$. The 
    }
\end{figure*}
After having extracted the strength of collisionless drag and found the regime where its presence is not negligible we study its effect in determining the behaviour of some correlation functions and its relationship with the Luttinger liquid (quantum hydrodynamic) low energy description of the two species Bose-Hubbard model.

Indeed in the 2SF phase, the low energy theory for our system corresponds to the Hamiltonian of two coupled Luttinger liquids \cite{giamarchi_book}. The Hamiltonian
can be diagonalised by introducing the density ($D$) and spin/polarisation ($S$) channels:
\begin{equation}
        H_{\mu} = \frac{1}{2\pi}\int \left[c_{\mu} K_{\mu}(\partial_x \phi_{\mu})^2 +\frac{c_{\mu}}{K_{\mu}}(\partial_x \theta_{\mu})^2 \right]\mathrm{d} x, 
        \label{eq::hydro_hamiltonian}
\end{equation}
where $\phi_{D(S)}=(\phi_{A}\pm\phi_{B})/\sqrt{2}$ and $\theta_{D(S)}=(\theta_{A}\pm\theta_{B})/\sqrt{2}$ are the bosonic fields related to the fluctuations of the phase and the amplitude of the total density (spin) of the two coupled superfluids \footnote{ In particular the long-wave length representation of the single particle operators are \begin{equation*}
    b_{x,\alpha}\rightarrow(n+\partial_x\theta_\alpha)^{1/2}\sum_m\exp[2\pi\textrm{i}m(\theta_\alpha+\pi n x)]e^{\textrm{i}\phi_\alpha}
\end{equation*}  see e.g. \cite{giamarchi_book,Clark}}. The speeds of sound $c_{D(S)}$, and $K_{D(S)}$ are the so-called Luttinger parameters.
There's an additional non-linear coupling between the densities of the two Luttinger liquids which can be perturbatively accounted for by a term proportional to ${U_{AB}\cos(2\sqrt{2}\theta_S)}$. This term is irrelevant in the 2SF phase and relevant in the PSF phase.
As long as the Hamiltonian Eq.~(\ref{eq::hydro_hamiltonian}) holds, an algebraic decay characterises the correlation functions (a.k.a. quasi-long-range order)~\cite{giamarchi_book}:
\begin{equation}
\begin{split}
        G_\alpha(x) &= \langle b^\dagger_{i+x,\alpha}b_{i,\alpha}\rangle \propto |d|^{-\frac{1}{4K_D}-\frac{1}{4K_S}},\\
        R_D(x) &= \langle b^\dagger_{i+x,A}b^\dagger_{i+x,B}b_{i,B}b_{i,A} \rangle \propto |d|^{-\frac{1}{K_D}},\\
        R_S(x) &= \langle b^\dagger_{i+x,A}b_{i+x,B}b^\dagger_{i,B}b_{i,A}\rangle \propto |d|^{-\frac{1}{K_S}}.
\end{split}
\label{eq:lutt_param}
\end{equation}
Here we expressed the algebraic decay in terms of the natural measure of the distances between sites on a ring geometry, i.e., the \textit{chord} function~\cite{Cazalilla2011}:
\begin{equation}
    d(x/L) = \frac{L}{\pi} \sin \left(\frac{\pi x}{L}\right) \, ,
\end{equation}
where $L$ is the number of sites and $x\in \mathbb{N}$ the linear distance between the sites. For very large rings the expression further simplifies according to the substitution $d\rightarrow x$.
The relations in \eqref{eq:lutt_param} can be easily checked by using the leading term in the long-wavelength field representation $b_{j,\alpha}\propto\exp{(i\phi_\alpha)}$~\cite{Haldane-hydro}.

The single-body correlations, $G_\alpha$, have a mixed density/spin character, consistently with the fact that the imaginary part of their nearest-neighbor value gives back the species current $j_\alpha$. 
The two contributions can be instead isolated with the help of two-body correlations: $R_D$ concerns the superfluid character of pairs of $A-B$ particles and therefore the density channel, while $R_S$ relates to particle-hole pairs and therefore the spin channel~\cite{Clark,Babaev2018}.

Away from commensurate effects, possibly leading to a Mott insulator, the density channel is always superfluid, i.e., $R_D$ scales algebraically.
A change in $R_S$ and $G_\alpha$ from algebraic to exponential decay -- or equivalently a drop of $K_S$ to 0 --  happens instead when entering the PSF phase, due to the opening of a gap in the spin channel.
This is illustrated in Fig.~\ref{fig:correlations}, where the correlations measured for different system sizes ($L=8, 16, 32, 64, 96$) are reported for two sample parameter values deep in the 2SF (blue) and PSF (orange) phases. \\

The Luttinger parameters satisfy the relation~\cite{giamarchi_book}:
\begin{equation}
    K_S = \pi\hbar \chi c_S/2,
\label{eq:lutt_sound}
\end{equation}
where  $\chi = \left[\partial^2 e / \partial (\nu_A-\nu_B)^2\right]^{-1}$ is the spin susceptibility %
with $e$ the energy density (the same goes for the density channel with the compressibility rather than the susceptibility).

On the other hand a hydrodynamic approach based on the energy functional including the collisionless drag~\cite{Andreev-Bashkin,Nespolo} provides the following relation between the spin-speed of sound and the superfluid densities%
~\footnote{Let us remind that the in a Galilean invariant systems one has that at $T=0$ the total superfluid density is equal to the total density, i.e., $2(n_{AA}^{(s)}+n_{AB}^{(s)})=n$, and therefore the numerator of Eq.~\ref{eq:hydro_sound} would be fully determined by the drag~\cite{Nespolo}.%
}:
\begin{equation}
    c_S^2 = 2\frac{n_{AA}^{(s)}-n_{AB}^{(s)}}{m^* \chi} \, .
\label{eq:hydro_sound}
\end{equation}
By direct comparison, the following also holds:
\begin{equation}
    K_S = \sqrt{\frac{\pi^2 \hbar^2 \, \left(n_{AA}^{(s)}-n_{AB}^{(s)}\right) \, \chi}{2m^*}} \, .  
    \label{eq:KS-hydro}
\end{equation}
The drag thus appears in the constitutive relations of the Luttinger parameters.
This fact, often overlooked in literature~\cite{Kollath1,Kollath2,Clark,Clark2,Citro2017,Citro2018}, is crucial in obtaining consistent results in the perturbative approach of Luttinger liquids.
Therefore in writing the two-species Luttinger Hamiltonian a term proportional to $n_{AB}^{(s)} \partial_x \phi_A \partial_x \phi_B$ must be included. The latter term should be indeed generated under renormalisation group flow obtained by means of the operator product expansion as it has been studied for the Josephson tunneling between superfluids (see, e.g.,~\cite{GiamarchiCastellano} and reference therein). 
\begin{figure}[tbh]    \centering
    \includegraphics[width = 0.48\textwidth]{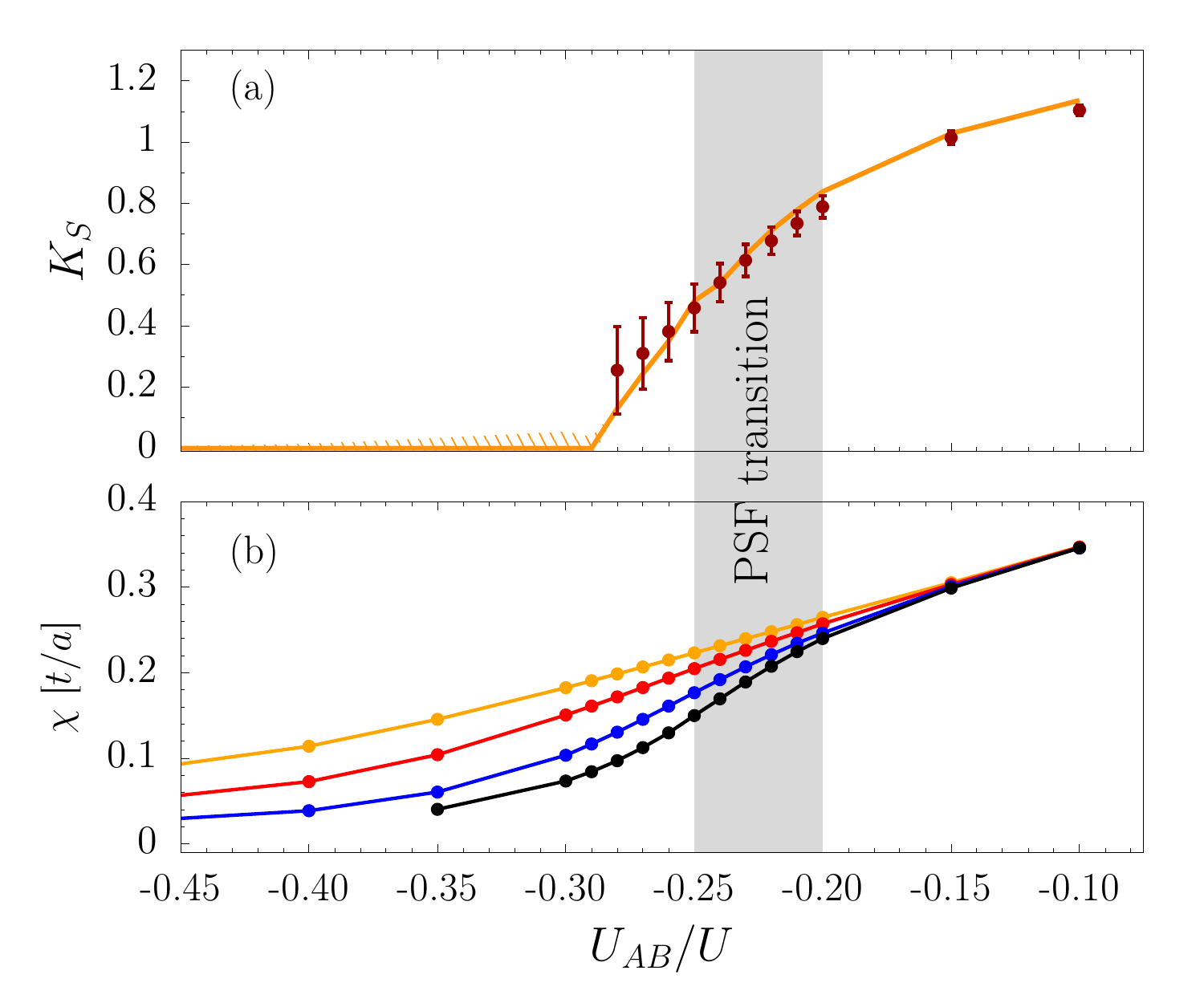}
    \caption{\label{fig:sound_speed}(color online). 
    In (a) Luttinger parameter $K_S$ for the spin channel, as obtained from the hydrodynamic relation in Eq.~\eqref{eq:KS-hydro} (solid line with error shadow) and from the correlation functions, Eq.~\eqref{eq:lutt_param} (points with error bars). 
    Points are reported until the algebraic fit makes sense: the shaded region indicates where deviations become sizeable (for more details see~\cite{SuppMat}). In the PSF the parameter $K_S$ must go to zero.
    (b) displays the behaviour of the susceptibility as function of the interaction, as estimated from various system sizes (color code as in Fig.~\ref{fig:dragPSF} except for $L=8$ that is omitted).
    }
\end{figure}
In Fig.~\ref{fig:sound_speed} we report the values of $K_S$ obtained by hydrodynamics (Eq.~\eqref{eq:KS-hydro}) and by the long-range behaviour of the correlation functions (Eq.~\eqref{eq:lutt_param}).

The latter, however, stops to be algebraic and becomes exponential once the system enters in the PSF phase~\cite{Clark}:
we quantify this change by measuring the deviation from a pure algebraic behaviour in log-log scale~(see~\cite{SuppMat} for a detailed discussion). The region where the deviation becomes appreciable is in agreement with the region predicted from the drag saturation (Fig.~\ref{fig:dragPSF}). Before the fluctuations region the two estimates for $K_S$ give consistent results. 

For the sake of comparison with mesoscopic experiments accessible with ultra-cold gases in optical lattices, we report in Fig.~\ref{fig:sound_speed} (b) also the estimated spin susceptibility for different system's sizes.
%


\section{Conclusions}
In conclusion, we provide a numerical estimation of the AB superfluid drag via a Tensor Network approach for a 1D Bose mixture on a ring lattice. The drag is enhanced for attractive interactions due to the relevance of paring correlations in 1D systems. In particular close to the 2SF-PSF transition, the entrainment can become of the same order of the full superfluid density, which makes this regime more suitable for an experimental measure. 
Differently from the Mott-Superfluid case for a single species~\cite{MatteoNJP,GiamarchiCastellano}, the thermodynamic estimate of the PSF transition proves to be challenging from a numerical point of view due to the slow convergence to the thermodynamic limit for the spin-channel~\cite{Parisi2018}. We stress again that our results are relevant for ultra-cold gases experiment where tens to a few hundreds of atoms are considered, where the drag could be extracted by measuring the susceptibility~\cite{Ferrari} and the spin speed of sound~\cite{Shin}.  


Moreover our analysis suggest that the inclusion of the drag has fundamental implications in the understanding of any hydrodynamic Luttinger liquid approach to the two species Bose-Hubbard model.
In particular we show that AB hydrodynamics provides a reliable expression for the Luttinger parameters of the spin-channel in terms of the superfluid densities and the susceptibility of the system. 

\textit{Note added:} while completing the present work we became aware of a DMRG-QMC comparison for the study of the pairing properties of one-dimensional hard-core bosons~\cite{Batrouni2020}. The results show the slow convergence to the thermodynamic limit for the spin-channel as in our soft-core case.
\section*{Acknowledgments}
We gratefully acknowledge insightful discussions with S. Giorgini, A. Haller, S. Manmana, J. Nespolo, and S. Stringari. M.R. acknowledges partial support from the
Deutsche Forschung Gesellschaft (DFG) through the individual Grant No. 277810020 (RI 2345/2-1) and the CRC network TRR183 Grant No. 277101999 (unit B01), the European Union (PASQuanS, Grant No. 817482), the Alexander von Humboldt Foundation, and the kind hospitality of the BEC Center in Trento, where a great part of this study was
performed. D.C. acknowledges hospitality at the Johannes Gutenberg University Mainz, where preliminary stages of this work were conducted. A.R. acknowledges financial support from the Italian MIUR under the PRIN2017 project CEnTraL,
the Provincia Autonoma di Trento, and the Q@TN initiative.
The MPS simulations were run on the Mogon Cluster of the Johannes Gutenberg-Universität Mainz (made available by the CSM and AHRP) and on the JURECA Cluster at the
Forschungszentrum Jülich, with a code based on a flexible Abelian Symmetric Tensor Networks Library, developed in collaboration with the group of S. Montangero (Padua).


%

\clearpage



\pagebreak

\makeatletter
\renewcommand{\theequation}{S\arabic{equation}}
\renewcommand{\bibnumfmt}[1]{[S#1]}
\renewcommand{\citenumfont}[1]{S#1}
\section*{SUPPLEMENTAL MATERIAL}
\setcounter{figure}{6}
\setcounter{page}{1}
\setcounter{section}{0}
\setcounter{equation}{0}
\setcounter{table}{0}


\section{Realization of periodic boundary conditions}
\label{app:numerics}
Periodic Boundary Conditions usually pose a challenge to Tensor Network methods due to two main reasons: the increased amount of entanglement to be accounted for when bi-partitioning the system in segments (twice the boundaries), and the presence of loops impairing the existence of a canonical form if naively mirrored in the network structure~\cite{MatteoAnthology}.
While the first is unavoidable, the latter could be circumvented by either employing Tree Tensor Networks~\cite{MatteoAnthology,MatteoNJP} or -- as we did here -- by mapping the ring onto a chain with tailored next-nearest neighbor couplings and boundary nearest neighbor ones~\cite{MatteoAndreasPRR}.

\begin{figure}[hbt]    \centering
    \includegraphics[width = 0.25\textwidth]{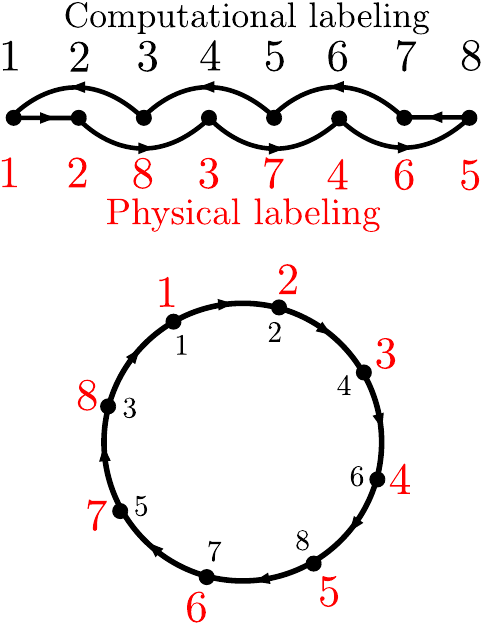}
    \caption{\label{fig::pbc}(color online). Sketch of the scheme used to mimic periodic boundary conditions with a computational open boundary conditions ansatz.}
\end{figure}

The loop therefore disappears from the MPS form and gets adsorbed inside the MPO. 
The basic idea of such ``snake'' pattern is shown in Fig.~\ref{fig::pbc} for a very little system with an even number of sites. 
Despite a slightly bigger MPO and a bit of extra book-keeping in the post-processing of observable measurements, the numerical calculations demonstrated to be efficient and stable~\cite{MatteoAndreasPRR}. 
%

\section{Numerical derivatives and errors}
\label{app:numerical_error}
The first source of error in the MPS ansatz that has been taken into account is the maximum local bosonic occupation number. The maximum total (species A and B) number of particles at site $x$ is chosen to be $N_{x}^{max}=6$. Because of the low filling of the computed systems, we checked that the probabilities of the high-occupation states beyond our truncation are negligible. 

\begin{figure}[hbt]    \centering
    \includegraphics[width = 0.48\textwidth]{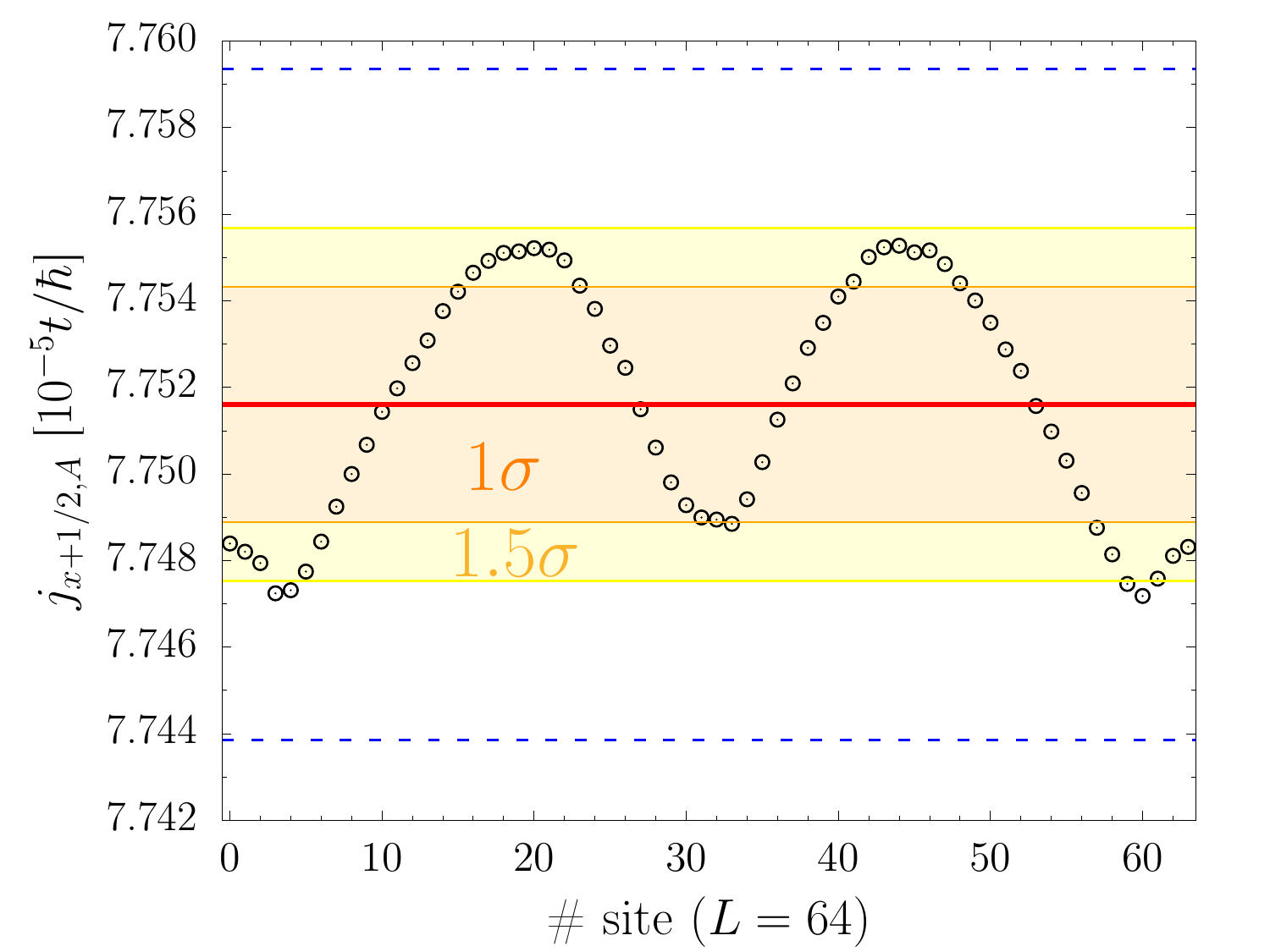}
    \caption{\label{fig:numerical_error}(color online). The $j_{x+a/2,A}$ current as function of $x$, indicated by the site labels. The system's configuration is: $t = 1$, $U = 10$, $U_{AB} = -3$, $\nu = 1$, $\phi_A = 0.002$ and $\phi_B = 0$. The red solid line is the average of the values and the shaded areas represent $1\sigma$ and $1.5\sigma$ error respectively. Notice that almost all the values are included in the $1\sigma$ region that is well inside the $0.1\%$ error (dashed blue line).}
\end{figure}

In order to compute the superfluid densities, we resorted to a four-point
numerical approximation of the derivative,
while the currents have been measured from the MPS wavefunction:
\begin{equation}
    \frac{\partial j_\alpha}{\partial \phi_\beta}\biggr\rvert_{0} \simeq 
    \frac{\bar{j_\alpha}[-2\delta_\beta]
    -8\bar{j_\alpha}[-\delta_\beta]
    +8\bar{j_\alpha}[\delta_\beta]
    -\bar{j_\alpha}[2\delta_\beta]}%
    {12 \delta_\beta} + O(\delta_\beta^4)
\label{eq::finitedifferencederivative}
\end{equation}
where the mean current of species $\alpha$ over the links of the lattice is $\bar{j_\alpha} = 1/L\sum_{x=1}^{L}j_{x+1/2,\alpha} $. 
The $O(\delta_\beta^4)$ error descending from the finite-difference derivative is negligible with respect to the one associated with the inhomogeneity of the currents $j_{x+1/2,\alpha}$ along the ring, shown in Fig.~\ref{fig:numerical_error}.
We have estimated the latter via the standard deviation $\sigma$ of the measured values and then propagated the uncertainty in Eq.~\eqref{eq::finitedifferencederivative}. Moreover, the resulting error has been compared and eventually combined with the error estimation due to the MPS ansatz. This has been accounted for by comparing the values of the final quantity at different bond dimensions' values. The whole procedure leads to the error bars reported in the figures of the main text.

Concerning the numerical error on the energy, we made use of the standard deviation of the ground state energy estimations along the last sweep of the MPS optimization (e.g., Chapter 5 of \cite{MatteoAnthology}).
This should be closely related to the accumulation of the truncated probabilities in the renormalization procedure.

Lastly, some presented quantities are the thermodynamic estimations unless explicitly specified. The errors result from the fit of those quantities as function of the inverse of the system's size $L$. More details on the extraction procedure of the Luttinger parameters from the correlation functions are provided in the following section.

%

\section{Extraction of the Luttinger parameters from the correlation functions}
\label{app:luttingerextraction}
\begin{figure}[hbt]    \centering
    \includegraphics[width = 0.48\textwidth]{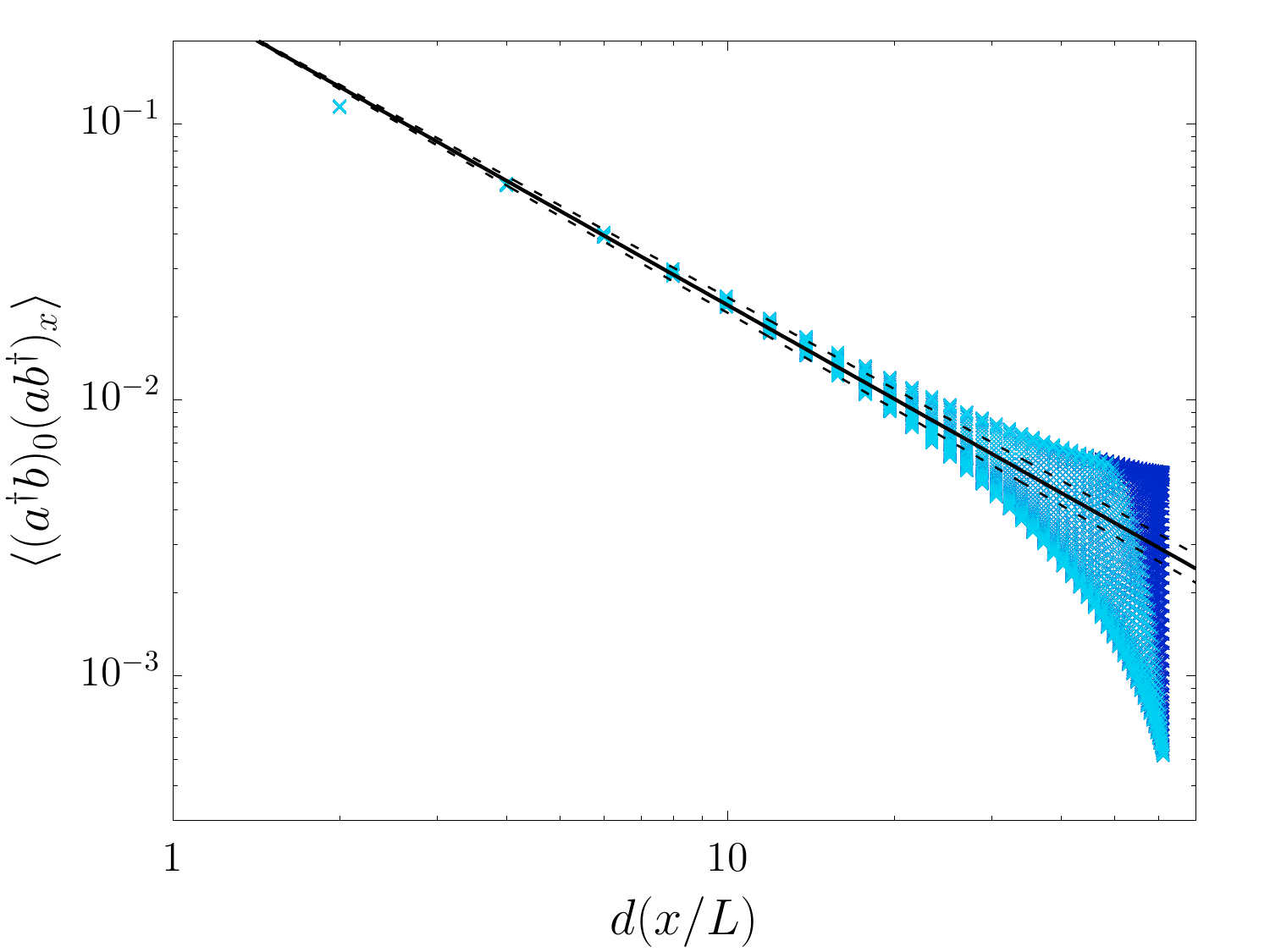}
    \caption{\label{fig::filtering}(color online). The effect of the filtering procedure for the $R_S$ correlations for a $L=96$ ring with $t = 1$, $U = 10t$, $U_{AB} = -0.2U$. The dark-blue points include all the correlations extracted from the MPS ansatz and the light-blue are the ones left from discarding those which are too far from a computational point of view thanks to the procedure explained in the main text. The algebraic fit is represented with a solid black line and the area between the two dashed black lines is the error on the fit. }
\end{figure}

In order to compute the Luttinger parameter $K_S$, we employed the following procedure:
first, we extracted at each finite size $L$ the exponents of the algebraic fits in Eq.~(9) and checked their mutual consistency. 
Then, we extrapolated the thermodynamic limit of $K_S(L)$ as function of $1/L$, as we also did for other quantities across the manuscript.

It is important to recall here that any MPS Ansatz leads to correlation functions expressed as a sum of exponential functions (as dictated by the underlying transfer matrix structure, see~\cite{SchollwoeckReview, OrusReview}), which could mimic even algebraic decays in finite systems. 
Due to our specific MPS structure of Fig.~\ref{fig::pbc}, however, the same physical distance might be represented by pairs of points at a pretty different computational distance.
Therefore, some of these pairs return more precise estimates than others: for this reason, when performing the (algebraic) fits, we filtered out data coming from pairs further than the half of the ring length ($x>L/2$) in the computational basis.
The result is shown in Fig.~\ref{fig::filtering}, where it is evident that the filtering allows for cleaner fits. 

\begin{figure}[hbt]    \centering
    \includegraphics[width = 0.48\textwidth]{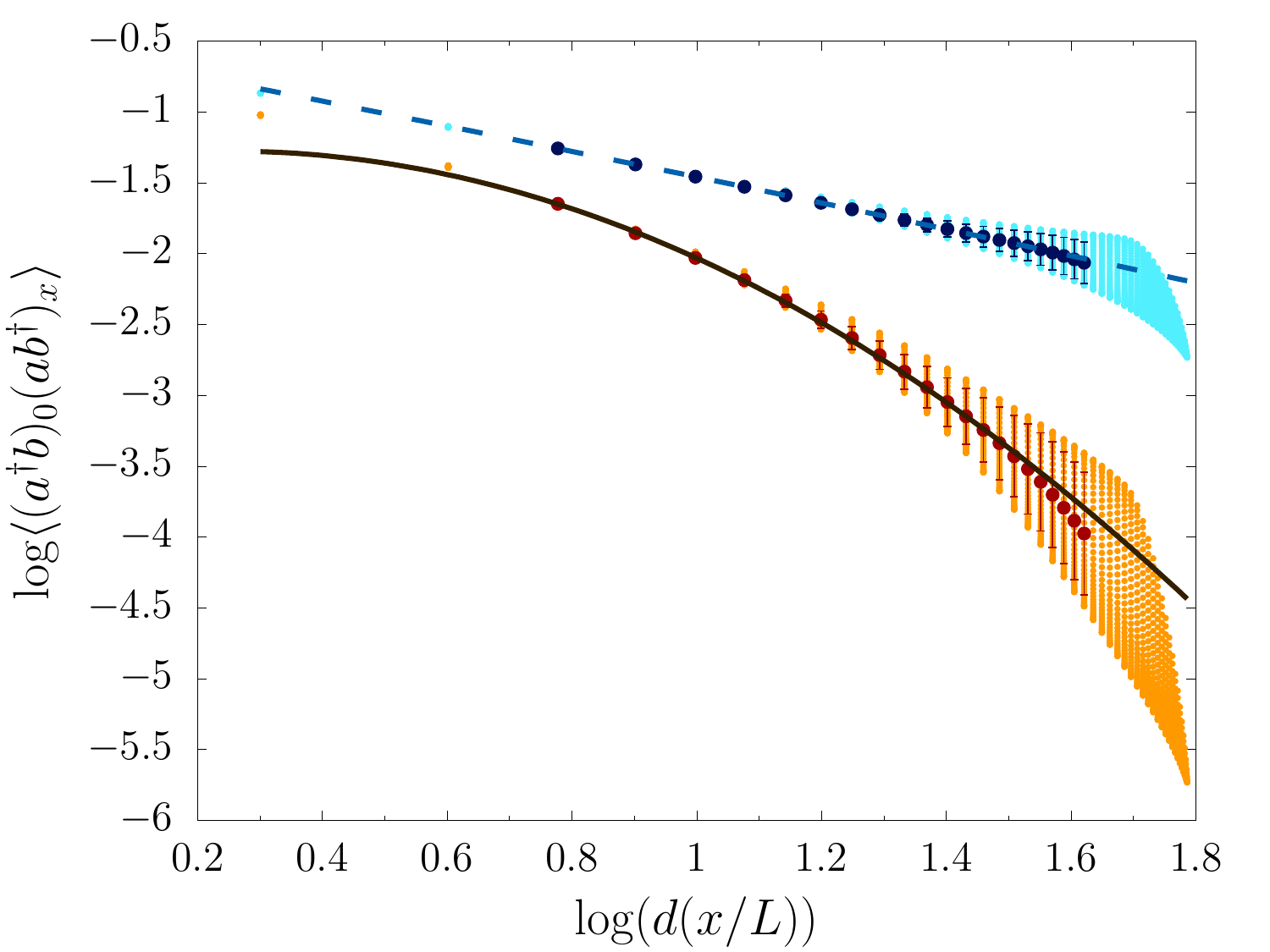}
    \caption{\label{fig::curvature}(color online). Fit of the correlations $\log[R_S(x)]$ as function of the logarithm of the conformal distance with a fitting function as Eq.~\eqref{eq::fit_with_alpha}. The configuration of the system is the same of Fig.~\ref{fig::filtering} except for the interpecies interactions: red-like point are for $U_{AB}/U=-0.3$ while blue-like are for $U_{AB}/U=-0.1$. The correspondent fitting functions (solid and dashed lines respectively) consider only the dark points with correspondent error bars. The bending of the correlations from a pure algebraic trend can be captured by increasing the $\alpha$ amplitude.}
\end{figure}

\begin{figure}[b]    \centering
    \includegraphics[width = 0.48\textwidth]{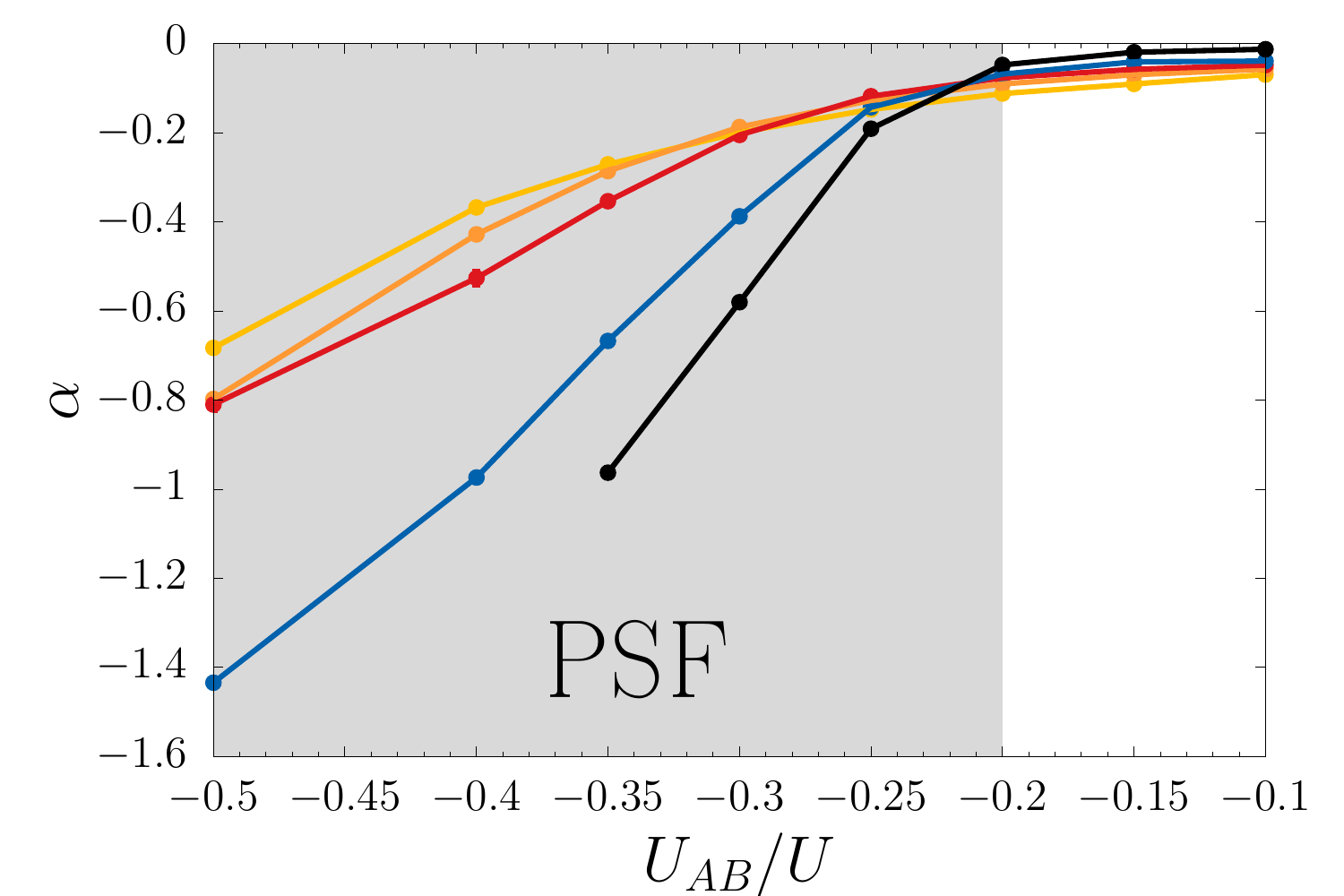}
    \caption{\label{fig::alpha}(color online). The values of $\alpha$ for different system's sizes. An increase in the $\alpha$ amplitude is associated to an exponential trend of the $R_S$ correlation function. }
\end{figure}

Finally, in order to better locate the 2SF to PSF transition region, we resorted as in~\cite{MatteoNJP} to an estimate of the curvature of the fitting function in the $\log$-$\log$ plane, quantified by $\alpha$:
\begin{equation}
    \log[R_S(x)] = \alpha \xi^2 - \frac{1}{K_S}\xi + \mathrm{const.}, 
    \label{eq::fit_with_alpha}
\end{equation}
where $\xi=\log[d(x/L)]$.
If $\alpha$ is zero the fit is purely algebraic and the system is in the 2SF phase (see dashed blue line in Fig.~\ref{fig::curvature}). Otherwise, if $\alpha$ assumes a finite value, the fit bends (down) from the algebraic trend and hence it signals an exponential kicking in.
This allows us in principle not to make any assumption on a critical value $K_S^*$ at which the phase transition happens.
In Fig.~\ref{fig::alpha} we show the $\alpha$ values as a function of the inter-species interaction $U_{AB}$, for various system sizes -- with the same color code used in the main text, from light to dark shades $L=8,16,32,64,96$ ($U/t=10$).
We can clearly spot two regimes: for $U_{AB}/U\geq-0.2$, $\alpha$ appears to converge towards zero for larger system sizes, while for $U_{AB}/U\leq-0.25$ the (negative) curvature $\alpha$ increases in amplitude with growing system sizes.
This latter behaviour hints at the fact that an algebraic fit of $R_S$ ceases to make sense beyond this region, which we therefore identify as the one where the 2SF-PSF transition takes place.

Larger system sizes and a more accurate sampling of the putative transition region would be needed in order to precisely locate the critical point -- e.g., the thermodynamic limit might emerge for a very large number of particles~\cite{Parisi2018}.
We leave such a more complete study of the transition for future work.
%


\end{document}